**Title: The recovery of George Berkeley's objective science of 1710 and its implications for traditional science.**

**Author: S. Clough (University of Nottingham, UK)**

In 1710, George Berkeley, philosopher and scientist [1], discovered that the directions in which we see (sightlines) are tilted 45 degrees towards the past from the directions in which we look. Looking north from the 3-d present we see to the north-past and so see the 4-d past. The past only contains images of the material objects in the present, so the world we see contains no material objects. We do not see a tree; we see an image of the tree, or, with two mirrors, an infinite stream of images of the tree receding into the 4-d past. The science of the 4-d past described with data-defined language is objective. When the Relativistic Doppler Effect is so described, it is simple and complete, unlike Einstein's space and time version in his 1905 theory of relativity. Scientists still use space and time science though, so Berkeley's discoveries have been neglected and lost. Their recovery in this paper reinstates the choice between a data-defined language description of the observed 4-d past and an idea-based language description of idea-based (3+1)-d space and time.

## 1. Science, ideas, language and data.

Since science began, people have proposed explanations in ordinary language for scientific data which can not be described in ordinary language. This has resulted in opinions being sought from scientists on the plausibility of explanations. The practice is not objective and does not comply with Occam's Razor. It also leads to ideas being accorded the status of reality.

Berkeley believed that science must be objective [2]. To achieve this, he aimed to exclude all ideas from science by replacing ordinary language by a new data-defined language. Words which are not data-defined are idea-defined so words like 'mountain', house and stream, are idea-defined and must be replaced.

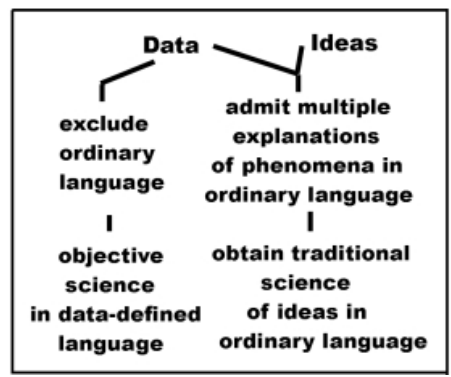

**Fig 1: Berkeley's route to objective science.**

Thus Berkeley planned to revolutionise science and his discovery that sightlines are tilted to the past made his plan practicable [3].

## II: Sightlines tilted to the past.

Fig 2 shows an observer B standing between a pair of mirrors facing each other (so-called infinity mirrors [4]) in a small room. He sees infinite lines of images of him and the mirrors stretching far away into the past. He also perceives that all the images are linked by an invisible line originating at his eyes and zig-zagging between the images of the mirrors in directions alternating between north-past (NP) and south-past (SP). When B looks into the southern mirror reflections in the mirrors cause his look lines to alternate between S, N, S etc while his sightline alternates between SP, NP, SP, NP, etc and goes deeper and deeper into the past.

The terms, 'sightline', 'image stream'. 'north-past', south-past', '4-d past' are data-defined. They do not occur in ordinary language.

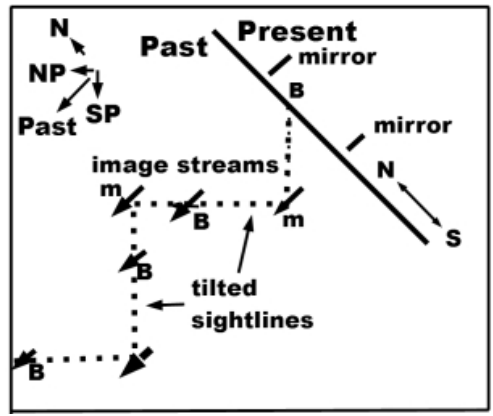

**Fig 2. Image streams in the past linked by a NP/SP sightline.**

In fig 3 a traditional scientist, T, sees exactly the same as B in fig 2, but interprets it very differently. He interprets the zig-zag line as a lightpath tilted by 45 degrees to the future by the advance of time. That leads to the traditional science of optics. In choosing fig 2 or fig 3 one is choosing between objective science in the past or traditional science in the present. This is the moment of decision for science.

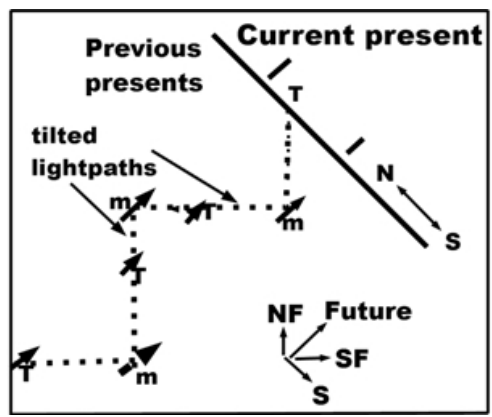

**Fig 3: The traditional science interpretation of fig 2 is of image streams in the advancing present linked by future-tilted lightpaths.**

Some spectacular examples f image streams can be found on the web [4].

## III Science in the 4-d past

Directions in the past are combinations of pairs of sightline vectors like NP and SP or NWP and SEP. Fig 4 compares the directions in a 2-d NWP/SEP plane of the 4-d past with the directions in a reference frame of traditional space and time science.

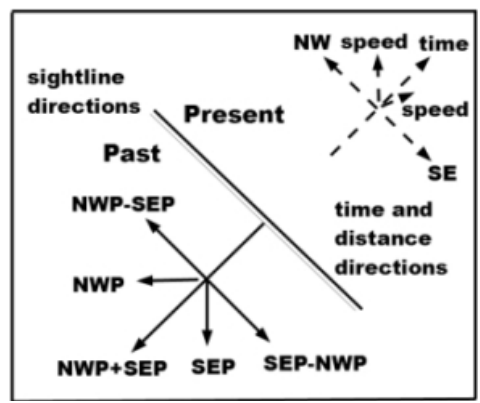

**Fig 4: A comparison of directions in objective science and traditional science.**

In fig 5, the coorginate grids are compared. The coordinate grids of objective science are data. Those of traditional science are ideas.

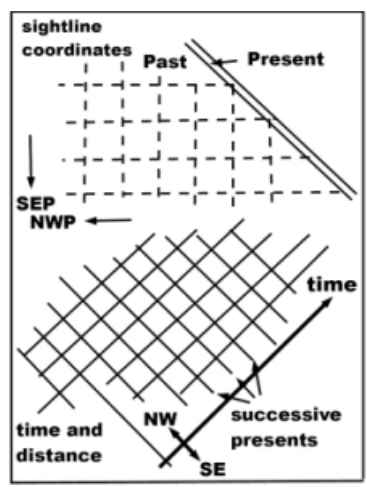

**Fig 5: A comparison of the coordinate grids of objective and traditional science.**

In objective science, objects in the present are continually changing their image and discarding their old image into the past. An object's image stream is thus a history of the object's progress through life. The past is a library of such histories.

Fig 6 shows how two people in the present see images of each other in the past at the point where the past-tilted sightline of one intercepts the other's image stream.

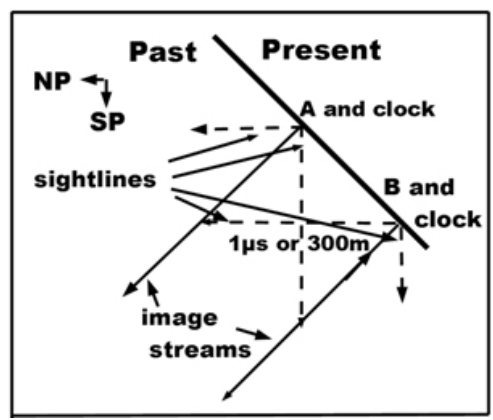

**Fig 6: How we inhabit the present but see the past.**

Because measurements in the past are all along sightlines, they can be made in either metres or seconds. It is found that 300 metres of sightline is equal to 1 μs of sightline. In traditional science this ratio, a number, appears as the speed of light, an idea, with metres of distance per microsecond of time, as a unit of speed.

Drivers steer in the present and feel the solidity of the steering wheel in the present, while watching images of the road ahead in the past. Surgeons hold the handle of a scalpel in the present while watching images of the blade in the past. Philosophers think in the present while being seen to think in the past. All movement is in the present. The past is a record of completed movements.

### IV: Examples of objective science in the 4-d past

When a pendulum weight in the present is still, the advance of its image stream into the past is along the direction (UP+DP) where UP=Up-Past and DP=Down-Past. The extent of the advance can be shown by adding a number of units, p, as in (UP+DP)(p). When the pendulum in the present swings along the NW/SE line the image stream advance is described by $(UP+DP)(p)+(NWP+SEP)q\sin(\omega p)$. There are no ideas of time, distance or speed in this description.

When Galileo dropped a musket ball from the Tower of Pisa, Galileo's image stream advanced along (UP+DP)(n), and the musket ball's image stream along $(UP+DP)(n)+(DP-UP)(gn^2/2)$.

The image of a bright star on a dark night is easy to see if there are no sightline-scattering clouds in the way. One simply looks up-past in the known direction and there it is. It is similarly easy for an electron detector to detect the image of an electron in the epr experiment. A sightline of an electron detector, filtered to detect electrons with a particular spin direction, is arranged to intersect an image stream of electron pairs with anti-parallel spin directions. When the filtered detector detects an electron image with its selected spin direction, an image of the electron detector containing the electron appears at the start of the detector's image stream.

In a radar measurement in the 4-d past. observer X begins by briefly illuminating the present to create a bright spot on his image stream which he sees later when it is intercepted by his sightline reflected from the target. From the image stream of a clock by his side, he learns the number of microseconds between him and the target.  As these preliminary examples show, the geometry and rules of objective science are simple and the vocabulary is limited to a few data-defined words.

# V: The objective science of the Relativistic Doppler Effect (RDE).

If observers Y and X are close together on the NW/SE line of the present and then move steadily apart, their image streams in the past diverge as shown in fig 5. If Y and X also carry mirrors to reflect a sightline to and fro between them, alternating in direction between NWP and SEP, the sightline is divided by its reflectios into segments of relative length, 1, 1/D, $1/D^2$, $1/D^3$ etc where D and $D^{-1}$ are the gradients of the image streams. The series, 1, 1/D, $1/D^2$, $1/D^3$, with the number D, is the signature of the relativistic Doppler Effect (RDE). The angle between the image streams is $\alpha_D$.

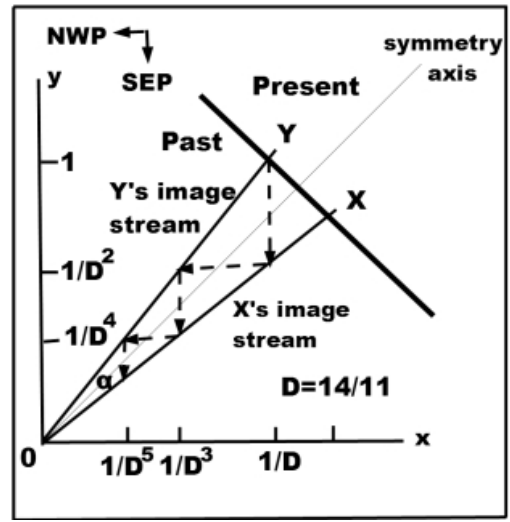

**Fig 7: Divergent image streams in the NWP/SEP plane of the past**

The parameters $S_D=\sin(\alpha_D)$ and $C_D=\cos(\alpha_D)$ are related to D by (1-5).

$$D^2=(1+S_D)/(1-S_D) \qquad (1)$$

$$S_D=(D-D^{-1})/(D+D^{-1}) \qquad (2)$$

$$C_D=2/(D+D^{-1}) \qquad (3)$$

$$C_D D = (1+S_D) \quad (4)$$

$$C_D/D = (1-S_D) \quad (5)$$

## VI: The unsymmetrical Relativistic Doppler Effect.

By making the ratio of the sizes of units on the x and y axes equal to σ=D/g the divergence of the image streams in fig 5 is made to appear unsymmetric. The gradients appear to be g and gD$^{-2}$ but the signature of the chain of arrows is unchanged at 1, 1/D, 1/D$^2$ 1/D$^3$. Figs 6 and 7 are examples.

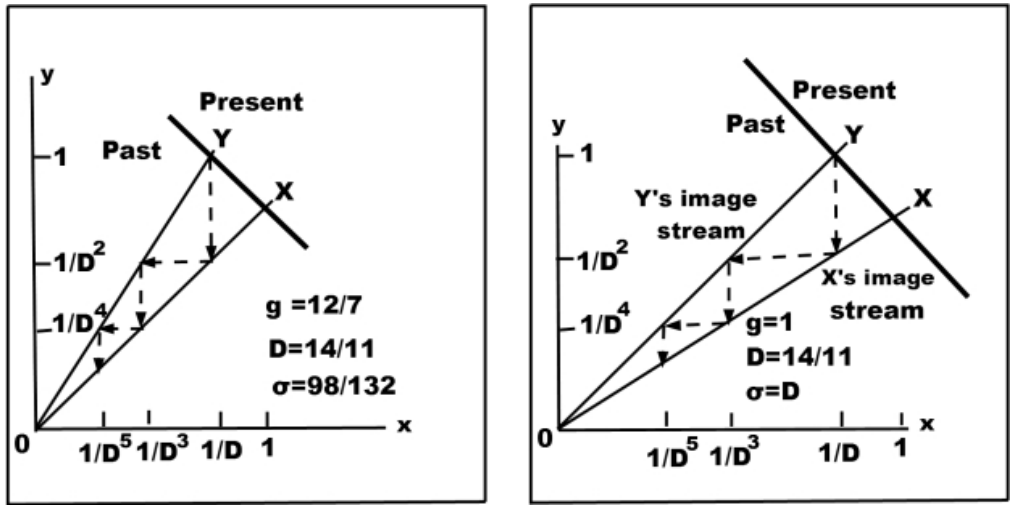

**Figs 8 and 9: Unsymmetric versions of fig 7.**

If a third image stream, Z's image stream, with gradient, 1, is added to diagrams like figs 5 and 6, three pairs of image streams, Y/X, Y/Z and Z/X are defined with $D_{YX}=D$, $D_{YZ}=g^{1/2}$ and $D_{ZX}=g^{-1/2}D$. These satisfy (6) and (7).

$$D_{YX}=D_{YZ}D_{ZX} \quad (6)$$

$$(1+S_{YX})/(1-S_{YX})=((1+S_{YZ})/(1-S_{YZ}))((1+S_{ZX})/(1-S_{ZX})) \quad (7)$$

Equation (7) may then be rearranged into (8) and (9).

$$S_{YX}=(S_{YZ}+S_{ZX})/(1+S_{YZ}S_{ZX}) \quad (8)$$

$$S_{YZ}=(S_{YX}-S_{ZX})/(1-S_{YX}S_{ZX}) \quad (9)$$

## VII: The impossibility of explaining data with traditional science

In traditional science, every number is attached to an idea like distance or time. In obective science numbers are attached to sightlines. All data are attached to sightlines so traditional science can not explain them. To deal with this problem, scientists call data 'raw data', attach ideas, and then call it 'data'. The basic diagram of the RDE, fig 7, is completely defined by one number or datum, which can be D or $S_D$ or $C_D$. Traditional science needs two independent variables, t and d, for the very different ideas of time and distance. Even if attaching ideas to data is accepted, it is impossible to attach two independent ideas to fig 7. Einstein's description of the RDE in his 1905 theory of relativity was therefore bound to fail. It fails most obviously in the example popularly known as the twin paradox.

Figs 11, 12 and 13 show how a symmetrical RDE diagram, fig 11, (like fig 7 but with D=2) can be made to resemble a traditional reference frame by first giving it the shape of a reference frame with an asymmetry parameter σ=D and then transforming to (τ, δ) coordinates with y=(τ-δ) and x=(τ+δ)/D where y=-SEP and x=-NWP are the original sightline coordinates. That is as far as one can go though. It is not possible to replace τ and δ by the ideas, time and distance because τ and δ are combinations of numbers attached to sightlines and only one is independent.

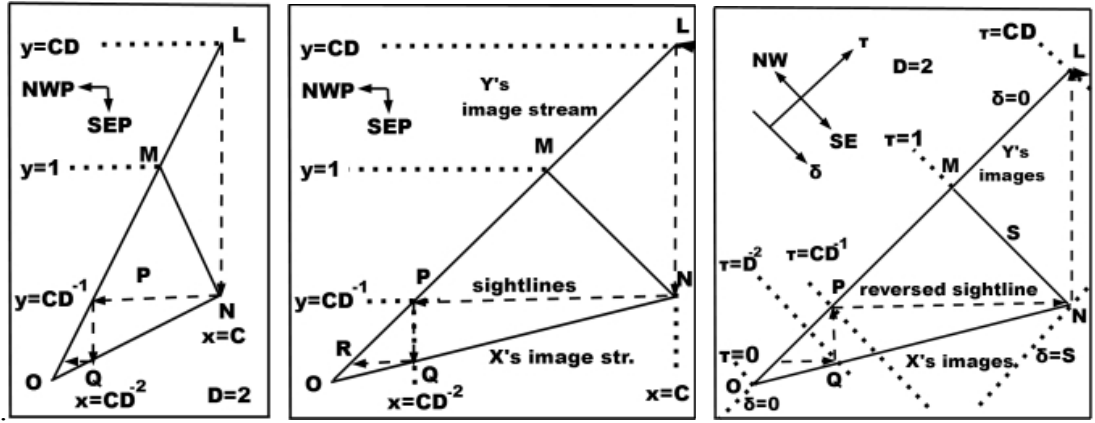

**Figs 10, 11 and 12. How an RDE diagram can be made to resemble a reference frame of traditional science.**

## VIII: Einstein's Theory of Relativity

In his theory, Einstein was trying to go from a reference frame like fig 12 to fig 10, encumbered by the ideas of distance and time which he wanted to take with him. He began by constructing a reference frame showing the paths of two diverging material objects through space and time and the paths of light passing to and fro between the objects. He then put the speed of light equal to a constant, so that the reference frame resembled fig 12. Then, instead of transforming to y and x to obtain fig 11, which would have meant losing the t and d coordinates, he invented the idea of time dilation which gave him the properties of fig 11 with the coordinates of fig 12. However, he also obtained the redundancy equation (10) telling him to change his coordinates from those of fig 12 to those of fig 11.

$$\gamma^1 = (1-(v/c)^2)^{1/2} \qquad (10)$$

He ignored this instruction though and retained his idea-based variables. Ostensibly, Einstein had converted a reference frame with the ideas of time and distance into a description of data-defined numbers. In reality, he had imposed the ideas of time and distance on the numbers and by ignoring the warning of the redundancy equation, had lost the ability to interpret the theory.

## IX: The twin paradox

Fig 13 shows two image streams which diverge from O and then converge to U.

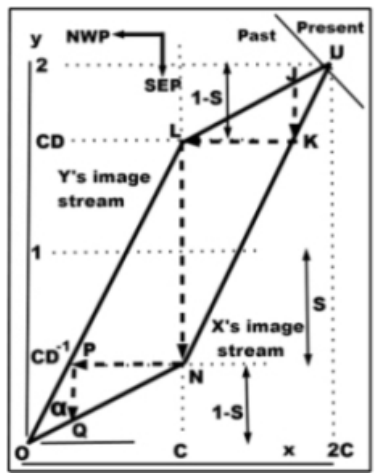

**Fig 13: The diagram known as the twin paradox.**

Y's clock measures the number of units that O is SEP of U and X's clock the number of units O is NWP of U. Y's clock records 2 units and X's clock 2C units. Einstein though, had left his version of the theory with time and distance coordinates so he interpreted fig 8 as the history of two twins who have the same ages at O and different ages at U due to time dilation. He was thus unaware of the real meaning of his theory.

When the variables of Einstein's theory are changed to S and C, all of his symbols collapse to one of the three numbers, 1, S or C and his equations $E=\gamma mc^2$, $p=\gamma mv$ and $E^2=p^2c^2+m^2c^4$ are meaningless redundancy equations. His conclusion that a relative speed can not exceed the speed of light reduces to the trivial fact that

sin($\alpha_D$) can not exceed 1 and what he identified as the curvature of light becomes the curvature of a sightline.

## X: Berkeley quotes.

The following Berkeley quotations are taken from ref 1.

*The sort of explanation proper to science then, is not causal explanation but reduction to regulariy'.*

That is Berkeley's claim to have found objective science.

. '*It is indeed an opinion strangely prevailing amongst men, that houses, mountains, rivers, and in a word all sensible objects have an existence natural or real, distinct from their being perceived by the understanding.*

That is Berkeley's exclusion of idea-defined words and language like houses etc from the description of the objective reality he had found in the 4-d past.

## Summary.

There can be little doubt that objective science is the science of the future. That, however, depends on scientists developing the data-defined language. There may be reluctance to give up the freedom to invent ideas like Einstein's constancy of the speed of light and time dilation, which scientists enjoy with traditional science.

Berkeley's reputation will be reassessed. He was not the maverick who insisted on believing something obviously absurd. He was a scientist who knew the importance of objectivity and set out to find it and succeeded.

The accessibility of infinity mirrors means that everybody can participate in a decisive event in the history of science. Science is at a fork. It must go forward to objectivity or backwards to invention. The decision should not be left to scientists alone.

**References.**